\documentclass[12pt]{iopart}
\usepackage{graphicx}

\usepackage{amsfonts}
\usepackage{amsthm}

\usepackage{setstack}

\catcode`@=11
\@addtoreset{equation}{section}

\catcode`@=12

\newtheorem{theorem}{Theorem}[section]

\newcommand{\nn}{\nonumber \\}

\newcommand{\diff}[2]{ \frac{\partial #1}{\partial #2} }

\begin{document}

\title{A directed walk model of a long chain polymer in a slit
with attractive walls
}
\author{R.~Brak$\dagger$, A.L.~Owczarek$\dagger$,
  A.~Rechnitzer$\dagger$  and S.G.~Whittington$\ddagger$ \\
  $\dagger$Department of Mathematics and Statistics,\\
  The University of Melbourne,\\Parkville, Victoria 3010, Australia\\
  $\ddagger$Department of Chemistry,\\
  University of Toronto,\\Toronto, Canada, M5S 3H6.
}

\begin{abstract}
  We present the exact solutions of various directed walk models of
  polymers confined to a slit and interacting with the walls of the
  slit via an attractive potential. We consider three geometric
  constraints on the ends of the polymer and concentrate on the long
  chain limit. Apart from the general interest in the effect of
  geometrical confinement this can be viewed as a two-dimensional
  model of steric stabilization and sensitized flocculation of
  colloidal dispersions. We demonstrate that the large width limit
  admits a phase diagram that is markedly \emph{different} from the
  one found in a half-plane geometry, even when the polymer is
  constrained to be fixed at both ends on one wall. We are not able to
  find a closed form solution for the free energy for finite width, at
  all values of the interaction parameters, but we can calculate the
  asymptotic behaviour for large widths everywhere in the phase plane.
  This allows us to find the force between the walls induced by the
  polymer and hence the regions of the plane where either steric
  stabilization or sensitized flocculation would occur.
\end{abstract}
 \bigskip

\noindent{\textbf{Short title:} A directed walk model of a long chain polymer in a slit
with attractive walls } 
\vfill
\maketitle

\section{Introduction}
\setcounter{equation}{0}

When a long linear polymer molecule in dilute solution is confined
between two parallel plates the polymer loses configurational entropy
and exerts a repulsive force on the confining plates.  If the monomers
are attracted to one of the two confining plates, the force is still
repulsive, but less is magnitude because the polymer can adsorb on one
plate and therefore does not extend as far into solution.  The effect
of the second confining plate is then less and the loss of
configurational entropy is less.  If the monomers are attracted to
both plates the net force can be attractive at large distances and
repulsive at smaller distances.  At large distances the energy term is
dominant while at smaller distances the entropy loss is dominant.
These phenomena are related to the stabilization of colloidal
dispersions by adsorbed polymers (steric stabilization) and the
destabilization when the polymer can adsorb on surfaces of different
colloidal particles (sensitized flocculation).

One would hope to be able to say something about this problem for a
relatively realistic model of a polymer in a good solvent such as
a self-avoiding walk and the
self-avoiding walk case has been considered by several
authors (see for instance Wall \emph{et al} 1977 and 1978,
Hammersley and Whittington 1985).
When the self-avoiding walk is simply
confined between two parallel lines or planes, and
does not otherwise interact with the confining lines
or planes, a number of results are
available.  Let $c_n(w)$ be the number
of self-avoiding walks on the simple cubic lattice $\mathbb{Z}^3$,
starting at the origin and confined to have the $z$-coordinate of each
vertex in the slab $0 \le z \le w$.  Then it is known that the limit
\begin{equation}
\lim_{n\to \infty }n^{-1} \log c_n(w) \equiv \kappa (w)
\end{equation}
exists.  If $c_n$ is the number of $n$-edge self-avoiding walks with
no geometric constraint then the connective
constant of the lattice is given by
\begin{equation}
\lim_{n\to \infty } n^{-1} \log c_n \equiv \kappa.
\end{equation}
It is known that
\begin{enumerate}
\item
$\kappa(w)$ is monotone increasing in $w$,
\item
$\lim_{w \to \infty} \kappa(w) = \kappa$, and
\item
$\kappa(w)$ is a concave function of $w$.
\end{enumerate}
In two dimensions, the exact values of $\kappa(w)$ are known for small
$w$ (Wall \emph{et al} 1977).   
When the walk interacts with the confining lines or planes
almost nothing is known rigorously but the problem has been studied by
exact enumeration methods (Middlemiss \emph{et al} 1976).

To make progress with the situation in which the walk interacts
with the confining lines or planes one has to turn to
simpler models.  In a classic paper DiMarzio and Rubin
(1971) considered a random walk model on a regular lattice where the
random walk is confined between two parallel lines or planes, and
interacts with one or both of the confining surfaces.  They see
both attractive and repulsive regimes in their model, though they confine
themselves to the case where either the polymer interacts with only
one surface or equally with both surfaces.

In this paper we study a directed version of the self-avoiding walk
model.  We consider directed self-avoiding walks on the square
lattice, confined between two lines ($y=0$ and $y=w$).  We consider
the cases where the walk starts and ends in $y=0$ (confined Dyck
paths, or \emph{loops}), starts in $y=0$ and ends in $y=w$
(\emph{bridges}) and starts in $y=0$ but has no condition on the other
endpoint beyond the geometrical constraint (\emph{tails}).  Our model
is related to that of DiMarzio and Rubin (1971) but we consider the
more general situation where the interaction with the two confining
lines or planes can be different.  The techniques which we use are
quite different from those of DiMarzio and Rubin.

We derive results for the behaviour of these three classes of directed
walks as a function of the interaction of the vertices with the two
confining lines, and the width $w$. In particular, we find the
generating functions exactly for each of the three models. We find the
free energy and hence the phase diagram in the limit of large slit
width for polymers much larger than the slit width. We demonstrate
that this phase diagram is \emph{not} the one obtained by considering
the limit of large slit width for any finite polymer. We calculate the
asymptotics for the free energy in wide slits and so deduce the forces
between the walls for large slit widths. This allows us to give a
``force diagram'' showing the regions in which the attractive, repulsive,
short-ranged and long-ranged forces act.

\section{The model}
\begin{figure}[ht!]
\begin{center}
\includegraphics[width=10cm]{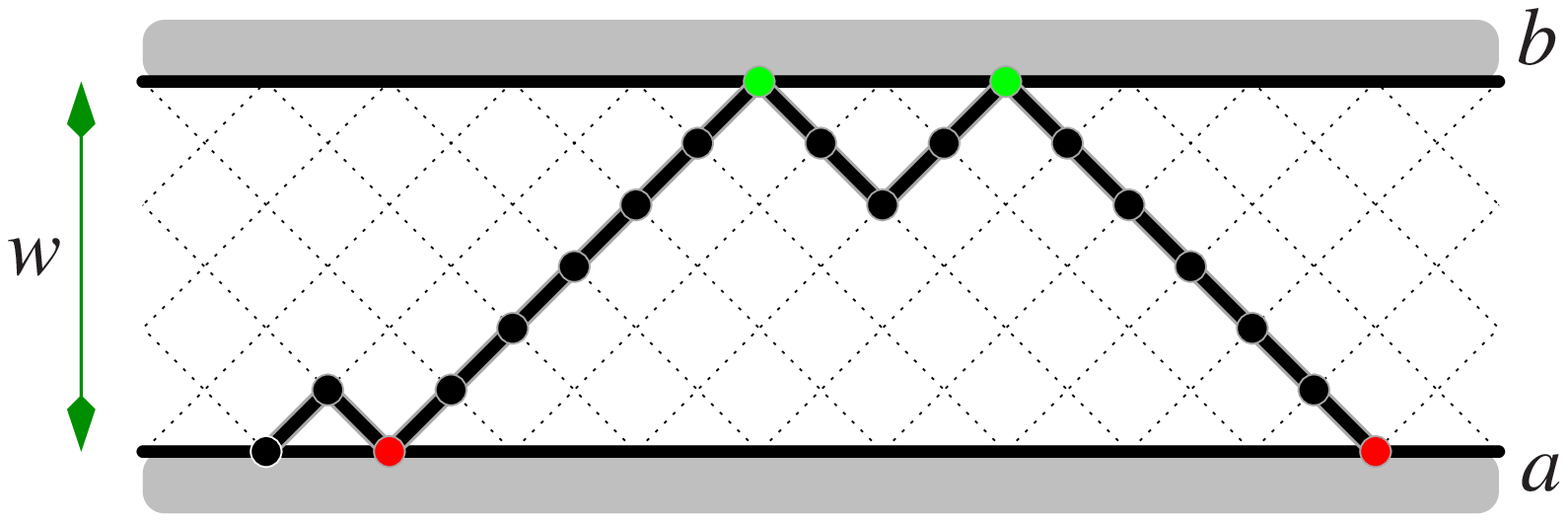}
\caption{\it  An example of a directed path which is a loop: both ends
of the walk are fixed to be on the bottom wall.}
\label{loop} 
\end{center}
\end{figure}

\begin{figure}[ht!]
\begin{center}
\includegraphics[width=10cm]{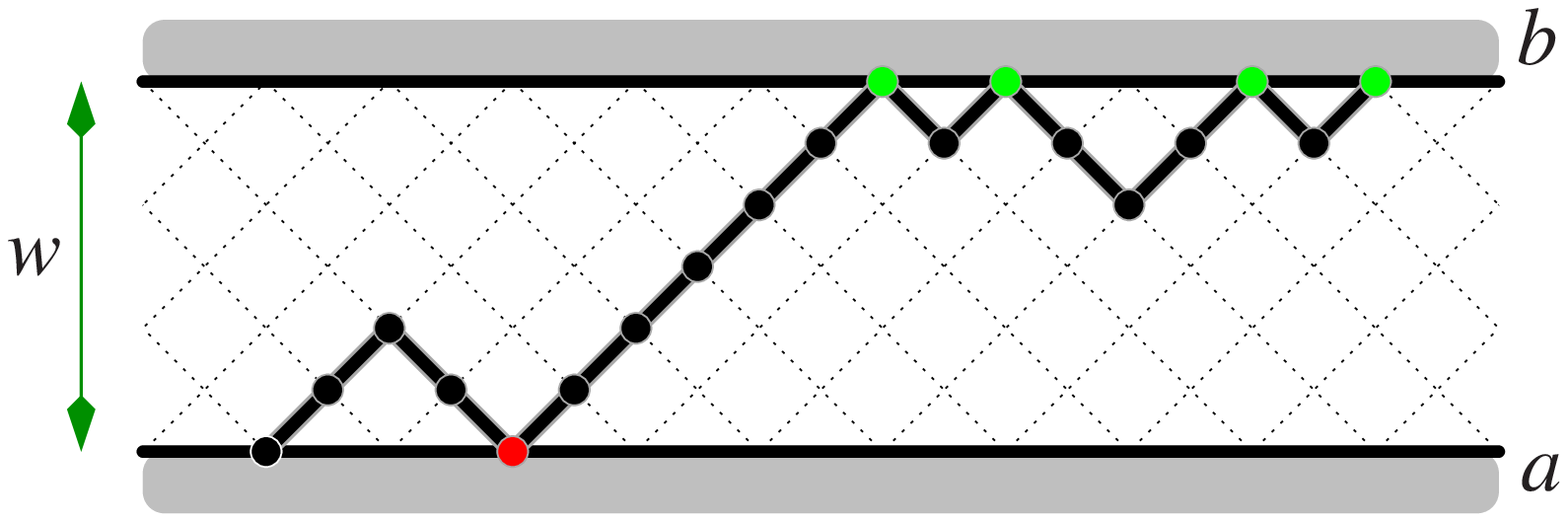}
\caption{\it  An example of a directed path which is a bridge: one end
of the walk is fixed to be on the bottom wall ($y=0$) while the other
end is fixed 
to be on the top wall ($y=w$).}
\label{bridge} 
\end{center}
\end{figure}

\begin{figure}[ht!]
\begin{center}
\includegraphics[width=10cm]{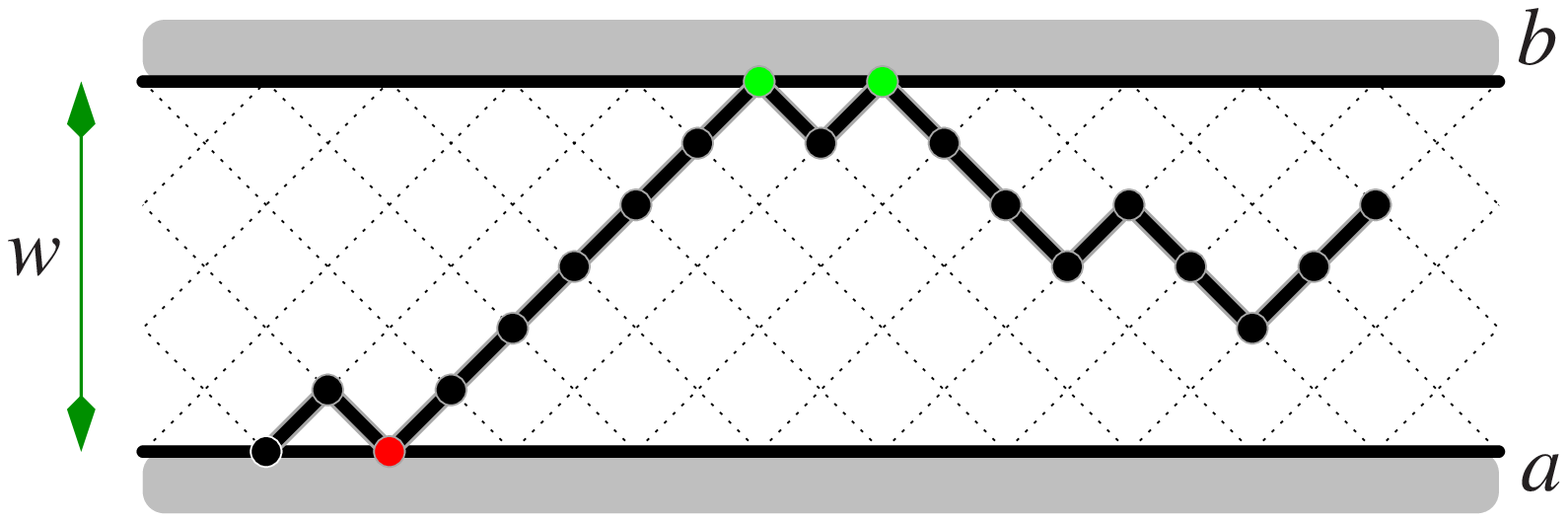}
\caption{\it  An example of a directed path which is a tail: one end
of the walk is fixed to be on the bottom wall ($y=0$) while the other
end  is not constrained.}
\label{tail} 
\end{center}
\end{figure}

The directed walk model that we consider is closely related to Dyck
paths (see Stanton and White 1986 or Deutsch 1999). These classical objects are closely related to Ballot paths which were  studied as a voting problem (Andre 1887 and  Bertrand 1887).     Dyck paths are directed walks on $\mathbb{Z}^2$ starting at
$(0,0)$ and ending on the line $y=0$, which have no vertices with negative
$y$-coordinates, and which have steps in the $(1,1)$ and $(1,-1)$
directions.  We impose the additional geometrical constraint that the
paths lie in the slit of width $w$ defined by the lines $y=0$ and
$y=w$. We refer to Dyck paths that satisfy this slit constraint as
\emph{loops} (see figure \ref{loop}). We also consider two similar
sets of paths; \emph{bridges} (see figure \ref{bridge}) and
\emph{tails} (see figure \ref{tail}). A bridge is a directed path lying in
the slit which has its zeroth vertex at $(0,0)$ and last vertex in
$y=w$. A tail is a directed path in the slit whose zeroth vertex is at
$(0,0)$ and whose last vertex may lie at any $0\leq y \leq w$.

Let $\mathcal{L}_w, \mathcal{B}_w$ and $\mathcal{T}_w$ be the sets of
loops, bridges and tails, respectively, in the slit of width $w$. We
define the generating functions of these paths as follows:
\begin{eqnarray}
  L_w(z,a,b) & = &  \sum_{p \in \mathcal{L}_w} z^{n(p)} a^{u(p)} b^{v(p)} ; \\
  B_w(z,a,b) & = &  \sum_{p \in \mathcal{B}_w} z^{n(p)} a^{u(p)} b^{v(p)} ;\\
  T_w(z,a,b) & = &  \sum_{p \in \mathcal{T}_w} z^{n(p)} a^{u(p)} b^{v(p)} ,
\end{eqnarray}
where $n(p), u(p)$ and $v(p)$ are the number of edges in the path $p$,
the number of vertices in the line $y=0$ (excluding the zeroth vertex)
and the number of vertices in the line $y=w$, respectively.

Let $\mathcal{L}_w^n, \mathcal{B}_w^n$ and $\mathcal{T}_w^n$ be the
sets of loops, bridges and tails of fixed length $n$ in the slit of
width $w$. The partition function of loops is defined as
\begin{equation}
Z_n^{loop}(w;a,b)  =   \sum_{p \in \mathcal{L}_w^n} a^{u(p)}
b^{v(p)}
\end{equation}
with the partition functions for bridges and tails defined analogously.
%\begin{eqnarray}
%  Z_n^{loop}(w;a,b) & = &  \sum_{p \in \mathcal{L}_w^n} a^{u(p)} b^{v(p)} \\
%  Z_n^{bridge}(w;a,b) & = &  \sum_{p \in \mathcal{B}_w^n} a^{u(p)} b^{v(p)} \\
%  Z_n^{tail}(w;a,b) & = &  \sum_{p \in \mathcal{T}_w^n} a^{u(p)} b^{v(p)} 
%\end{eqnarray}
Hence the generating functions are related to the partition functions
in the standard way, eg for loops we have
\begin{equation}
  L_w(z,a,b)  =   \sum_{n} z^{n} Z_n^{loop}(w;a,b) .
\end{equation}

We define the reduced free energy $\kappa^{loop}(w;a,b)$ for loops for
fixed finite $w$ as
\begin{equation}
\kappa^{loop}(w;a,b)  =  \lim_{n\rightarrow\infty} n^{-1} 
\log Z_n^{loop}(w;a,b)
\end{equation}
with the reduced free energies for bridges $\kappa^{bridge}(w;a,b)$
and tails $\kappa^{tail}(w;a,b)$ defined analogously.
%\begin{eqnarray}
%\kappa^{loop}(w;a,b) & = & \lim_{n\rightarrow\infty} n^{-1} 
%\log Z_n^{loop}(w;a,b)\\
%\kappa^{bridge}(w;a,b) & = & \lim_{n\rightarrow\infty} n^{-1} 
%\log Z_n^{bridge}(w;a,b) \\
%\kappa^{tail}(w;a,b) & = & \lim_{n\rightarrow\infty} n^{-1} 
%\log Z_n^{tail}(w;a,b).
%\end{eqnarray}

Consider the singularities $z_c^{loop}(w;a,b)$, $z_c^{bridge}(w;a,b)$
and $z_c^{tail}(w;a,b)$ of the generating functions $L_w(z,a,b)$,
$B_w(z,a,b)$ and $T_w(z,a,b)$, respectively, closest to the origin and
on the positive real axis, known as the critical points.  Given that
the radii of convergence of the generating functions are finite, which
we shall demonstrate, and since the partition functions are positive,
the critical points exist and are equal in value to the radii of
convergence.  Hence the free energies exist and one can relate the
critical points to the free energies for each $type$ of configuration,
(that is loops, bridges and tails) as
\begin{eqnarray}
\kappa^{type}(w;a,b)  =  -\log z_c^{type}(w;a,b).
\end{eqnarray}

\begin{theorem}
\label{samefe}
Loops, bridges and tails have the same limiting free energy at every
fixed finite $w$. 
\end{theorem}
\proof We sketch the proof: Fix $w$ at some finite positive
integer, and let $Z_n^{loop},
Z_n^{bridge}$ and $Z_n^{tail}$ be the partition functions of loops,
bridges and tails with $n$ edges in the strip of width $w$. By
appending $w$ edges to loops and bridges we have the inequalities
\begin{equation}
  b Z_n^{loop} \leq Z_{n+w}^{bridge} \qquad \mbox{ and } 
  \qquad a Z_n^{bridge} \leq Z_{n+w}^{loop}.
\end{equation}
Hence $ab Z_n^{loop}  \leq a Z_{n+w}^{bridge} \leq
Z_{n+2w}^{loop}$. Taking logarithms, dividing by $n$ and taking the
limit $n \to \infty$ shows that the limiting free energies are equal
(since the limits exist).

Since every loop is a tail $Z_n^{loop} \leq Z_n^{tail}$, and by
appending $w$ (if $n$ and $w$ have the same parity) or $w - 1$ edges
(if $n$ and $w$ have different parity) to tails we have
\begin{equation}
  a Z_n^{tail} \leq Z_{n+w}^{loop} \qquad \mbox{ or } 
  \qquad a Z_n^{tail} \leq Z_{n+w - 1}^{loop},
\end{equation}
depending on the parity of $n+w$. By a similar sandwiching argument
the limiting free energies of loops and tails, and hence bridges, are
all equal. \qed

Since the free energies are the same for the three models we
define $\kappa(w;a,b)$ as
\begin{equation}
\kappa(w;a,b) \equiv \kappa^{loop}(w;a,b) = \kappa^{bridge}(w;a,b)=
\kappa^{tail}(w;a,b) 
\end{equation}
with $z_c(w;a,b)$ being 
\begin{equation}
z_c(w;a,b) \equiv z_c^{loop}(w;a,b) = z_c^{bridge}(w;a,b)=
z_c^{tail}(w;a,b) 
\end{equation}
so that
\begin{equation}
\kappa(w;a,b) =  -\log z_c(w;a,b).
\end{equation}

\section{The half-plane}
%\subsection{Definitions}

For loops and tails we first consider the case in which $w \to
\infty$, which reduces the problem to the adsorption of paths to a
wall in the half-plane. We begin by noting that once $w > n$ for any
finite length walk there can no longer be any visits to the top
surface so
\begin{equation}
\mathcal{L}_{w+1}^n  =  \mathcal{L}_w^n \quad 
\mbox{ and } \quad 
\mathcal{T}_{w+1}^n  =  \mathcal{T}_w^n \mbox{ for all } w > n. 
\end{equation}
Hence we define the sets $\mathcal{L}_{hp}^n$ and
$\mathcal{T}_{hp}^n$ as
\begin{equation}
\mathcal{L}_{hp}^n  =  \mathcal{L}_{n+1}^n \quad 
\mbox{ and } \quad 
\mathcal{T}_{hp}^n  =  \mathcal{T}_{n+1}^n.
\end{equation}
The limit $w \to \infty$ can be therefore be taken explicitly.
%\begin{equation}
%\mathcal{L}_{hp}^n  =  \lim_{w\rightarrow\infty} \mathcal{L}_w^n \quad 
%\mbox{ and } \quad 
%\mathcal{T}_{hp}^n  =   \lim_{w\rightarrow\infty} \mathcal{T}_w^n.
%\end{equation}
Also, as a consequence of the above, for all $w > n$ we have $v(p)=0$
for any $p\in\mathcal{L}_w^n$ and also for any $p\in\mathcal{T}_w^n$.
Hence the partition function of loops in the half-plane can be defined as
\begin{equation}
  Z_n^{loop,hp}(a)  =   \sum_{p \in \mathcal{L}_{hp}^n} a^{u(p)} 
\end{equation}
with the partition function of tails in the half-plane defined analogously.
We define the generating function of loops in the half
plane via the partition function as
\begin{equation}
  L(z,a)  = \sum_{n} z^{n} Z_n^{loop,hp}(a) 
\end{equation}
with the generating function of tails $T(z,a)$ in the half-plane
defined analogously.  Note that the above limit does not exist for
bridges since for any fixed walk of length $n$ we have $w\leq n$.  In
an analogous way to the slit we define the reduced free energy in
the half-plane for loops (and tails analogously) as
\begin{equation}
\kappa^{loop,hp}(w;a)  =  \lim_{n\rightarrow\infty} n^{-1} 
\log Z_n^{loop,hp}(w;a) .
\end{equation}
We note that in defining these free energies for the half-plane the
thermodynamic limit $n\rightarrow\infty$ is taken
\emph{after} the limit $w \to \infty$: we shall return to this order
of limits later.

%\subsection{Summary of the half-plane solution}

While the half-plane solution is well known (Brak  \emph{et al} 1998,
2001 or  Janse van Rensburg 2000) it is useful to summarise the main results
for comparison with our new results for slits. One may factor both loops
and tails in the half plane by considering the second last point at
which a loop touches the line $y=0$ or the last point at which a tail
touches the surface.  This leads to the functional equations $L(z,a) =
1 + a z^2 L(z,a) L(z,1)$ and $T(z,a) = L(z,a) + z L(z,a) T(z,1)$ for
the generating functions for loops and tails.
%(see
%figure \ref{halfplanefactorl} and
%\ref{halfplanefactort})
%\begin{eqnarray}
%  L(z,a) & = & 1 + a z^2 L(z,a) L(z,1) \\
%  T(z,a) & = & L(z,a) + z L(z,a) T(z,1).
%\end{eqnarray}
%\begin{figure}[ht!]
%\begin{center}
%\includegraphics[width=14cm]{figures/dyck_fact.eps}
%\caption{\it  Half-plane factorization for loops: any loop is either a 
%single vertex or has the structure illustrated.}
%\label{halfplanefactorl} 
%\end{center}
%\end{figure}
%\begin{figure}[ht!]
%\begin{center}
%\includegraphics[width=14cm]{figures/tail_fact.eps}
%\caption{\it  Half-plane factorization for tails: any tail is either a 
%loop or has the structure illustrated.}
%\label{halfplanefactort} 
%\end{center}
%\end{figure}
These may be easily solved to give
\begin{eqnarray}
  \label{eqn gf L hp}
  L(z,a) & = & \frac{2}{2-a+a\sqrt{1-4z^2}} \quad \mbox{ and} \\
  \label{eqn gf T hp}
  T(z,a) & = & \frac{4z}{(a-2-a\sqrt{1-4z^2})(1-2z-\sqrt{1-4z^2})}.
\end{eqnarray}
The generating functions are singular when the square root term is
zero and when the denominator is zero. These conditions give the
locations of the critical points, $z_c^{loop,hp}(a)$ and
$z_c^{tail,hp}(a)$ and so also give the free energies. The values are
equal for the two models and so we define $z_c^{hp}(a)\equiv
z_c^{loop,hp}(a) = z_c^{tail,hp}(a)$, and similarly 
%\begin{equation}
%  z_c^{loop,hp} = z_c^{tail,hp}(a) = \left\{ \begin{array}{cc}
%    \frac{1}{2} & a\leq 2 \\
%    \frac{\sqrt{a-1}}{a} & a > 2
%  \end{array} \right.
%\end{equation}
%Defining the reduced free energy in the half-plane models analogously to the
%finite width cases 
for the free energy we have
\begin{equation}
\kappa^{hp}(a) \equiv \kappa^{loop,hp}(a) = 
\kappa^{tail,hp}(a) = -\log z_c^{hp}(a)
\end{equation}
with 
\begin{equation}
  \kappa^{hp}(a)= \left\{ \begin{array}{cc}
    \log(2) & a\leq 2 \\
    \log\left(\frac{a}{\sqrt{a-1}}\right) & a > 2 .
  \end{array} \right.
\label{halfplanefe}
\end{equation}
%which is illustrated in figure \ref{halfplanefe-diag}.
%\begin{figure}[ht!]
%\begin{center} 
%\includegraphics[width=10cm]{figures/zc_ads.eps}
%\caption{\it  A plot of the thermodynamic limit free energy for the
%half-plane limit. For $a\leq 2$ the free energy is constant
%while for $a>2$ it varies. \emph{I would prefer to plot 
%against log a since the free energy is a convex function of log a.}}
%\label{halfplanefe-diag} 
%\end{center}
%\end{figure}
The second derivative of $\kappa^{hp}(a)$ is discontinuous at $a=2$
and so there is a second-order phase transition at this point.
The density of visits in the thermodynamic limit, defined as
\begin{equation}
\rho^{hp}(a) = \lim_{n\rightarrow \infty} \frac{\langle v\rangle}{n} = \frac{d \kappa^{hp}(a)}{d \log{a}}
\end{equation}
is an order parameter for the transition and has corresponding exponent 
$\beta=1$.
%given in figure \ref{rho-hp}.
%\begin{figure}[ht!]
%\begin{center}
%\includegraphics[width=8cm]{figures/visit_density.eps}
%\caption{\it  The $a$-dependence of the limiting density of visits per step.
%This function plays the role of an order parameter. } 
%\label{rho-hp}
%\end{center}
%\end{figure}

%For fixed $a$, as $z \to z_c(a)^-$ the generating functions behave as
%$(z_c(a)-z)^{-\gamma(a)}$, where $\gamma(a)$ is the entropic critical
%exponent. By Darboux's Theorem we have
%\begin{equation}
%Z_n \sim A(a) e^{\kappa(a) n} n^{\gamma(a) -1} \mbox{ as } n \rightarrow \infty
%\end{equation}
%where $A(a)$ is independent of $n$.
%The value of the entropic exponent is different for the two models:
%\begin{equation}
%  \gamma_{loop,hp}(a) = \left\{ \begin{array}{cc}
%      -\frac{1}{2} & a < 2 \\
%      \frac{1}{2} & a = 2 \\
%      1 & a > 2
%    \end{array} \right.
%  \qquad
%  \gamma_{tail,hp}(a) = \left\{ \begin{array}{cc}
%      \frac{1}{2} & a < 2 \\
%      1 & a = 2 \\
%      1 & a > 2
%    \end{array} \right.
%\end{equation}

%%%%%%%%
%%%%%%%%

\section{Exact solution for the generating functions in finite width
strips}

\begin{figure}[ht!]
\begin{center}
\includegraphics[width=10cm]{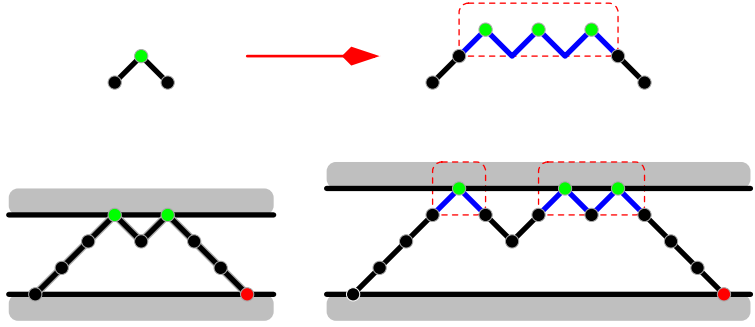}
\caption{\it  The construction of configurations in strip width $w+1$ from 
configurations in strip width $w$ is illustrated. Every vertex
touching the upper wall in the strip of width $w$ can be replaced by a 
\emph{zig-zag} path as shown.}
\label{rowbyrow} 
\end{center}
\end{figure}

The solution of the strip problem proceeds in an altogether different
manner to the half-plane situation.  This begins by using an argument
that builds up configurations (uniquely) in a strip of width $w+1$
from configurations in a strip of width $w$. In this way a
recurrence-functional equation is constructed rather than a simple
functional equation. Consider configurations of any type --- loops,
bridges or tails --- in a strip of width $w$ (see figure
\ref{rowbyrow}), and focus on the vertices touching the top wall: call
these \emph{top} vertices. These vertices contribute a factor $b$ to
the Boltzmann weight of the configuration. Now consider a zig-zag path
(see figure \ref{rowbyrow}), which is defined as a path of any even
length, or one of length zero, in a strip of width 1. The generating
function of zig-zag paths is $1/(1-bz^2)$. Replace each of the
\emph{top} vertices in the configuration by any zig-zag path. Since
one could choose a single vertex as the zig-zag path all configurations
that fit in a strip of width $w$ are reproduced. Also, the addition of
any non-zero length path at any top vertex will result in a new
configuration of width $w+1$ and no more. The inverse process is also
well defined and so we can write recurrence-functional equations for
each of the generating functions.
The generating function $L_w(z,a,b)$ satisfies the following
functional recurrence:
  \begin{eqnarray}
    L_1(z,a,b) & = & \frac{1}{1-abz^2} ;\\
    L_{w+1}(z,a,b) & = & L_w\left(z,a,\frac{1}{1-bz^2} \right).
        \label{loop-rec}  
\end{eqnarray}
The generating function of bridges, $B_w(z,a,b)$, satisfies a similar
functional  recurrence:
  \begin{eqnarray}
    B_1(z,a,b) & = & \frac{bz}{1-abz^2} ;  \\
    B_{w+1}(z,a,b) & = & bz \; B_w\left(z,a,\frac{1}{1-bz^2} \right).
\label{bridge-rec}  
  \end{eqnarray}
We note that the zeroth vertex of the path is weighted $1$ and that
$L_w(z,a,b)$ counts the walk consisting of a single vertex.
To compute the generating function of tails we introduce a new
variable $c$ such that a tail whose last vertex is in the line $y=w$
is weighted by $c$. Using this variable we arrive at the following
functional recurrence:
  \begin{eqnarray}
    \tilde{T}_1(z,a,b,c) & = & \frac{(1+cz)}{1-abz^2} ;\\
    \tilde{T}_w(z,a,b,c) & = & \tilde{T}_w\left(z,a,\frac{1}{1-bz^2},\frac{1+cz}{1-bz^2}\right).
\label{tail-rec}  
  \end{eqnarray}
The relevant tail generating function is then $T_w(z,a,b)=\tilde{T}_w(z,a,b,b)$.

Using these results it is easy to prove by induction that the
generating functions for any finite $w$ must be ratios of polynomials
in $z$. In fact using induction it is possible to prove that the
generating functions $L_w(z,a,b)$, $B_w(z,a,b)$ and $\tilde{T}_w(z,a,b,c)$ 
have the following forms:
  \begin{eqnarray}
    L_w(z,a,b) & = &  \frac{P_{w}(z,0,b) }{ P_{w}(z,a,b) },\\
    B_w(z,a,b) & = & \frac{b z^w }{ P_{w}(z,a,b) } ,\\
    \tilde{T}_w(z,a,b,c) & = & \frac{Q_w(z,b,c)}{P_w(z,a,b)},
        \label{gen-polyrat}
  \end{eqnarray}
  where the $P_w(z,a,b)$ and $Q_w(z,b,c)$ are polynomials.  By simply
  iterating the recurrence-functional equations and using, for example
  the GFUN package of MAPLE${}^{TM}$ or by combinatorial means
  (Flajolet 1980 and Viennot 1985), one can find the recurrences
  $P_{w+1} = P_w - z^2P_{w-1}$ and %
  $Q_{w+1} = (1+z)Q_w - z(1+z)Q_{w-1}+ z^3Q_{w-2}$. %
  The solution of these, using yet another generating
  variable $t$ over width $w$, gives
  \begin{equation}
  \fl \qquad
    G_p(z,a,b,t)\equiv \sum_{w\geq1} P_w(z,a,b) t^w = 
    t \times \frac{(1-abz^2) + (ab-a-b) z^2 t }{1-t+z^2t^2}
        \label{genfunforpw}
  \end{equation}
and
  \begin{equation}
  \fl \qquad
    G_q(z,a,b,c,t)\equiv\sum_{w\geq1} Q_w(z,b,c) t^w = 
    t \times \frac{ (1+cz) - zt(c+bz) + cz^3t^2}{(1-zt)(1-t+z^2t^2) }.
  \end{equation}
Alternately, one can find from substitution into the
functional-recurrences (\ref{loop-rec}) and (\ref{tail-rec}) that 
\begin{equation}
P_{w+1}(z,a,b)= (1-bz^2) P_w(z,a,\frac{1}{1-bz^2}) 
\end{equation}
and 
\begin{equation}
Q_{w+1}(z,b,c)= (1-bz^2) Q_w(z,\frac{1}{1-bz^2},\frac{1+cz}{1-bz^2}).
\end{equation}
By multiplying both sides by $t^w$ and summing over $w$, one obtains
functional equations for $G_p$ and $G_q$, which presumably can be
solved directly. We note that $P_w$ and $Q_w$ are polynomials related
to Chebyshev polynomials of the second kind (see for instance
Abramowitz and Stegun 1972 or Szeg\"{o} 1975) as can be seen by
comparing generating functions. Other techniques such as transfer
matrices (Brak \emph{et al} 1999), constant term (Brak \emph{et al} 1998) and
heaps (Bousquet-M\'elou and Rechnitzer 2002) would also undoubtedly
work on these problems.

\section{Location of the generating functions singularities }

\subsection{Transformation of the generating variable}
We recall that the limiting free energy of a model is determined by
the dominant singularities of its generating function. Since the
generating functions of these three models are all rational with the
same denominator, $P_w(z,a,b)$, it is the smallest real positive zero
of this polynomial that determines the limiting free energy. Theorem
\ref{samefe} shows that this zero (since the only zero of the
numerator for bridges is $0$ itself) is not cancelled by a zero of the
numerator. Below we find the limiting free energy by studying the
zeros of $P_w(z,a,b)$. Given $w$ we would like to find an expression
for $z_c$ as a function of $a,b$.

Since the generating function $G_p(t)$ for the $P_w$ polynomials is a
rational function of $t$ we can write it in partial fraction form with
respect to $t$, and then expand each simple rational piece to find
$P_w$ as a function of $w$.
%\begin{equation}
%  \sum_{w\geq1} P_w(z,a,b) t^w 
%  = t \times \frac{(1-abz^2) + (ab-a-b) z^2 t }{1-t+z^2t^2}.
%\end{equation}
The decomposition of the right-hand side of (\ref{genfunforpw})
involves expressions containing surds which becomes quite messy. It is
easier first to make a substitution that greatly simplifies the
partial fraction decomposition and the subsequent analysis. This is
chosen so as to simplify the denominator of $G_p(t)$. If we set
\begin{equation}
  z = \frac{\sqrt{q}}{1+q}
\label{zq}
\end{equation}
the generating function of the polynomials then simplifies to
\begin{eqnarray}
\fl \qquad
G_p(\sqrt{q}/(1+q),a,b, t) 
% \sum_{w \geq 1} P_w(\sqrt{q}/(1+q),a,b) t^w 
%  & = & t \frac{1+2q+q^2-abq+abqt-taq-tbq}{(1+q-t)(1+q-tq)} \\
  & = & ab-a-b+\frac{(1+q-aq)(1+q-bq)}{(1-q)(1+q-t)} \nn
  && - \frac{(1+q-a)(1+q-b)}{(1-q)(1+q-tq)}
\label{partfractq}
\end{eqnarray}
where there are simply linear factors in the denominator. The partial
fraction expansion has been taken in (\ref{partfractq}).  In our
discussion of the singularities of $P_w$ we will start with this
$q$-form. We can convert our results back later by making the inverse
substitution
\begin{equation}
  q = \frac{1-2z^2-\sqrt{1-4z^2}}{2z^2}.
\end{equation}

The above partial fraction form implies that
%  $P_w(\sqrt{q}/(1+q),a,b)$, with $w>0$ as
  \begin{equation}
    \fl \quad
    P_w(\sqrt{q}/(1+q),a,b) = 
    \frac{(1+q-aq)(1+q-bq)}{(1-q)(1+q)^{w+1}}   
    - \frac{(1+q-a)(1+q-b) q^w}{(1-q)(1+q)^{w+1}},
  \end{equation}
and when $a=0$ this simplifies to
  \begin{equation}
    \fl \quad
    P_w(\sqrt{q}/(1+q),0,b) = 
    \frac{(1+q-bq)}{(1-q)(1+q)^{w}}   - \frac{(1+q-b) q^w}{(1-q)(1+q)^{w}}.
  \end{equation}
We also have
%  \begin{equation}
%    \fl \quad
%    Q_w(\sqrt{q}/(1+q),a,b,b) = andrew 
%  \end{equation}
%\begin{eqnarray}
%\fl \qquad
%Q_w(\sqrt{q}/(1+q),a,b,c)&=&\frac{q^{\frac{1}{2} + w}\,\left( 1 - b + q
%\right) }
%   {\left( -1 + {\sqrt{q}} \right) \,\left( -1 + q \right) \,{\left( 1
%+ q \right) }^w} \nonumber \\
%&-& \frac{-1 - q + b\,q}
%   {\left( -1 + {\sqrt{q}} \right) \,\left( -1 + q \right) \,{\left( 1
%+ q \right) }^w} \nonumber\\
%   &+ &
%  \frac{q^{\frac{w}{2}}\,\left( -1 + c + \left( b - 2\,c \right)
%\,{\sqrt{q}} +
%       \left( -1 + c \right) \,q \right) }{{\left( -1 + {\sqrt{q}}
%\right) }^2\,
%     {\left( 1 + q \right) }^w}.
%\end{eqnarray}
\begin{eqnarray}
 \fl \qquad
Q_w(\sqrt{q}/(1+q),a,b,c)=&\frac{q^{\frac{1}{2} + w}\,\left( 1 - b + q
\right) }
   {\left(  1 - {\sqrt{q}} \right) \,\left( 1 - q \right) \,{\left( 1 +
q \right) }^w} \nonumber \\
&+
  \frac{1 +q - b\,q}
   {\left( 1 - {\sqrt{q}} \right) \,\left( 1 - q \right) \,{\left( 1 +
q \right) }^w} \nonumber \\
   &-
  \frac{q^{\frac{w}{2}}\,\left( 1 - c - \left( b - 2\,c \right)
\,{\sqrt{q}} +
       \left( 1 -c \right) \,q \right) }{{\left( 1 - {\sqrt{q}} \right)
}^2\,
     {\left( 1 + q \right) }^w}.
\end{eqnarray}

We can therefore find the generating functions for loops, bridges and
tails. The generating function $L_w(\sqrt{q}/(1+q),a,b)$ can be written as
  \begin{equation}
    \fl \qquad
    L_w   
%    & = &\frac{ P_{w}(\sqrt{q}/(1+q),0,b) }{
%P_{w}(\sqrt{q}/(1+q),a,b) }\nonumber\\ 
     =  \frac{(1+q) \left[(1+q-bq) - (1+q-b)q^{w}\right]}
    {(1+q-aq)(1+q-bq)-(1+q-a)(1+q-b)q^w}.
  \end{equation}
Similarly $B_w$ and $T_w$ can be written as
  \begin{equation}
    \fl \qquad
    B_w 
%    & = & \frac{b q^{w/2}}{(1+q)^w P_{w}(\sqrt{q}/(1+q),a,b)}\nonumber\\
     =  \frac{b q^{w/2}(1-q)(1+q)}{(1+q-aq)(1+q-bq)-(1+q-a)(1+q-b)q^w}
  \end{equation}
and   
\begin{equation}
\fl \qquad
T_{w}= \frac{\left( 1 + q \right) \,\left(
q^{\frac{1 + w}{2}} -1\right) \,
    \left(  q^{\frac{w}{2}}\,\left( 1 - b + q \right)   + \left( b-1
\right) \,q     -1    \right) }
    {\left(   {\sqrt{q}-1} \right) \,\left( q^w\,\left(   a-1 - q
\right) \,
       \left( b-1 - q \right)  - \left(   \left(   a-1 \right) \,q
-1\right) \,
       \left(   \left(   b-1 \right) \,q-1 \right)  \right) }.
\end{equation}

\subsection{Limit $w \to \infty$ of the generating functions}

In the limit that $w \to \infty$ we have, for $|q|<1$,  
\begin{eqnarray}
   & L_w(\sqrt{q}/(1+q),a,b) & \to   \frac{(1+q)}{1+q-aq}, \\
  & T_w(\sqrt{q}/(1+q),a,b) & \to 
  \frac{(1+q)}{(1+\sqrt{q})(1+q-aq)}  \\
  \mbox{and }\quad & B_w(\sqrt{q}/(1+q),a,b) & \to  0.
\end{eqnarray}
We note that substituting $z=z(q)$ back into the limits of $L_w$ and
$T_w$ gives the generating functions of loops and tails adsorbing in a
half-plane (equations~\ref{eqn gf L hp} and~\ref{eqn gf T hp}) as
required.

\subsection{Singularities of the generating function of loops}

Since each of the models has the same denominator polynomial this
confirms that all three models have the same limiting free energy.  To
determine the free energy we examine the dominant singularity of the
denominator polynomial.

The dominant singularity $q_c$ is the zero of the denominator of
$L_w(q/(1+q),a,b)$ that has the minimal value of $|z|$, and hence is
the solution of
\begin{equation}
q^w =
\frac{(1+q-aq)(1+q-bq)}{(1+q-a)(1+q-b)}  \
\label{q-sing-eqn}
\end{equation}
with this property.

\begin{figure}[ht!]
\begin{center}
\includegraphics[width=10cm]{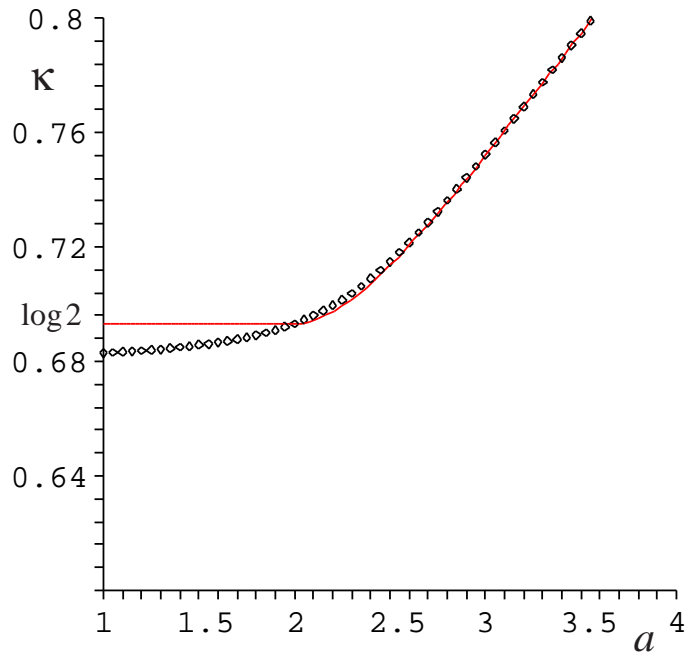}
\caption{\it  A plot of the free energy for width $w=20$ when $a=b$ is 
shown as the diamond markers. The solid curve is the half-plane
solution. }
\label{kappa-a=b} 
\end{center}
\end{figure}

\begin{figure}[ht!]
\begin{center}
\includegraphics[width=10cm]{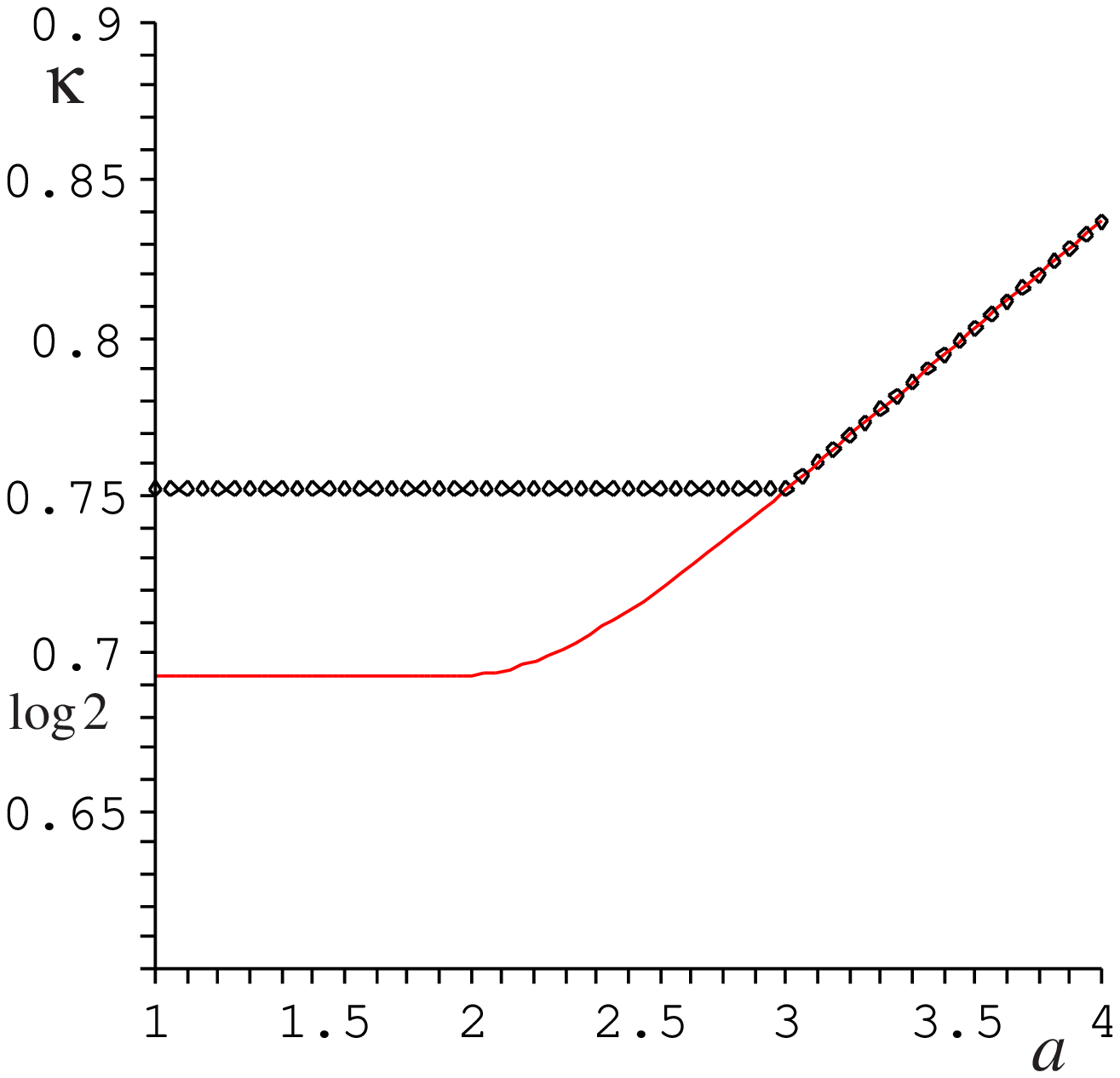}
\caption{\it  A plot of the free energy for width $w=20$ when $b=3$ is 
shown as the diamond markers. The solid curve is the half-plane
solution.}
\label{kappa-b=3} 
\end{center}
\end{figure}

Before discussing the cases in which we can find the singularity analytically,
we first consider some examples where we locate the singularity
numerically. To begin let us consider the symmetric case $a=b$ as this 
was also considered by DiMarzio and Rubin (1971).
In figure \ref{kappa-a=b} we plot $\kappa^{loop}(20;a,a)$ with the
half-plane thermodynamic limit for comparison.
For larger widths it is clear that the free energy converges to the
half-plane curve as one might expect.  Let us now consider weighting
the vertices in the top wall such that they would strongly attract
vertices of the polymer -- one may expect this to hamper the
convergence to the half-plane curve. Hence we consider $b=3$ and again
for width $w=20$ plot the free energy $ \kappa^{loop}(20;a,3)$:
this can be found in figure \ref{kappa-b=3}. 
It can now be seen that for $a<3$ the free energy is converging to
something close to $0.75$ and not the half-plane curve which is
$\log(2)$ for $a\leq2$. Physically, we can understand that a
long polymer in a slit with a highly attractive top wall will stick to
that wall while the attraction of the bottom wall is lesser in
magnitude. However, this points to the fact that this large width limit
for infinitely long polymers does \emph{not} demonstrate the same
physics as the half-plane case.

\section{Special parameter values of the generating function}
The solutions of the equation (\ref{q-sing-eqn}) cannot be written in
closed form for all values of $a$ and $b$. However, there are
particular values for which this equation simplifies. When $a=1$ or $2$ and 
$b=1$ or $2$, and when 
\begin{equation}
1-a = \frac{1}{1-b} \qquad \mbox{ equivalently } \qquad  ab=a+b  
\end{equation}
the right-hand side of equation (\ref{q-sing-eqn}) simplifies.
In these cases the locations of all the singularities of the
generating functions can be expressed in closed form for all widths $w$.
It is worth considering the generating function directly as
some cancellations occur at these special points.

For the special values mentioned 
the generating functions $L_w$ and
$B_w$ have particularly
simple forms, while the forms for $T_w$ are slightly more
complicated. We give the explicit form for $L_w$ 
at each of the points below:
  \begin{itemize}
  \item when $a=b=1$
    \begin{equation}
      L_w = \frac{(1+q)(1-q^{w+1})}{1-q^{w+2}};
 %     \qquad
%      B_w = \frac{(1-q)(1+q) q^{w/2} }{1-q^{w+2}};
    \end{equation}
  \item when $a=2$ and $b=1$
    \begin{equation}
      L_w = \frac{(1+q)(1-q^{w+1})}{(1-q)(1+q^{w+1})};
%      \qquad
%      B_w = \frac{(1+q)q^{w/2}}{1+q^{w+1}};
    \end{equation}
  \item when $a=1$ and $b=2$
    \begin{equation}
      L_w = \frac{(1+q)(1+q^{w})}{1+q^{w+1}};
 %     \qquad
 %     B_w = \frac{2(1+q)q^{w/2}}{1+q^{w+1}};
    \end{equation}
  \item when $a=b=2$
    \begin{equation}
      L_w = \frac{(1+q)(1+q^{w})}{(1-q)(1-q^{w})};
%      \qquad 
%      B_w = \frac{2(1+q)q^{w/2}}{(1-q)(1-q^w)}.
    \end{equation}
\item
on the curve $ab=a+b$   
\begin{equation}
L_w  =  \frac{(1+q)( (1+q-a) + 
        (1+q-aq)q^w)}{(1+q-a)(1+q-aq)(1-q^{w})} .
\end{equation}

%    \begin{eqnarray}
%      L_w & = & \frac{(1+q)( (1+q-a) +
%        (1+q-aq)q^w)}{(1+q-a)(1+q-aq)(1-q^{w})}  \\ 
%      B_w  & = & - \frac{a(1+q)(1-q)q^{w/2}}{(1-q^w)(1+q-a)(1+q-aq)}.
%    \end{eqnarray}
  \end{itemize}
%Similar, but slightly more complicated expressions may be found for
%$T_w(\sqrt{q}/(1+q),a,b)$ at these points.

The singularities in $z$ of $L_w$, $B_w$ and $T_w$ may be found from
the singularities in $q$ by the mapping (\ref{zq}). The singularities of
$L_w$ in $q$ occur either for $|q|=1$ or for $q \in \mathbb{R}^+$. The 
transformation (\ref{zq}) maps $q \in \{ |q|=1 \} \cup \mathbb{R}^+$ to $z
\in \mathbb{R}^+$. We note that if $q = e^{i \theta}$ then 
$1/z = 2 \cos(\theta/2)$. We concentrate on the dominant singularity
$z_c$ for which $|z|$ is minimal.

Using the expressions for $L_w$  given above we find the
zeros of the denominators (that do not cancel with a zero of the
numerator) and choose the one that minimises $|z|$. This gives us $q_c$
and hence $z_c$. From these we find the free energy $\kappa$ at the special
points listed above:
%taking account of cancellations in numerator and
%denominator (for example $q=1$ is not a singularity at $a=b=1$):
  \begin{itemize}
  \item when $a=b=1$ the zero is $q_c = \exp(2\pi i /(w+2) )$ and hence
    \begin{equation}     
     \kappa(w;1,1) = \log\left( 2 \cos(\pi/(w+2)) \right)    ;
    \end{equation}
  \item when $a=1$, $b=2$ and when $a=2$, $b=1$ the zero is
$q_c = \exp(\pi i / (w+1) )$ and hence
    \begin{equation}
      \kappa(w;2,1) = \kappa(w;1,2)= \log\left( 2 \cos(\pi/2 (w+1))\right);     
    \end{equation}
  \item when $a=b=2$ the zero is $q_c = 1$ and hence
    \begin{equation} 
     \kappa(w;2,2)= \log(2);
    \end{equation}
  \item when $ab=a+b$ and $a\geq b$ the zero is $q_c = (a-1)^{-1}$ 
        while when $ab=a+b$ and $b\geq a$ the zero 
        $q_c = (b-1)^{-1}=(a-1)$. Hence for $a>1$
    \begin{equation}
      \kappa(w;a,a/(a-1))  = \log\left(\frac{a}{\sqrt{a-1}}\right).
\label{spec-free-energy}
    \end{equation}
  \end{itemize}
In a similar way, one may write down expressions for the locations of all the 
singularities of the generating functions.

It is instructive to consider the free energy (see equation
(\ref{spec-free-energy})) along the special curve given by
$ab=a+b$. We first note that it is \emph{independent} of the width of
the slit. It is also clearly different at all values of $a<2$ to the
half-plane free energy (which is $\log(2)$). This re-iterates that the
large width limit of the slit problem is not necessarily described by
the half-plane results.
%\begin{figure}[ht!]
%\begin{center}
%\includegraphics[width=10cm]{figures/zc_f0_curve.eps}
%\caption{\it  Free energy of a slit of any width when $ab=a+b$.}
%\label{zeroforcefe} 
%\end{center}
%\end{figure}

In order to discuss the effective force between the walls for large
$w$ we need to calculate the large $w$ asymptotics of the free
energy. Since we have written down the closed form solution for all
$w$ of $\kappa(w;a,b)$ at some special points it is advantageous to
consider the large $w$ asymptotics of these first. As noticed above,
on the curve $ab=a+b$ the $\kappa$ is independent of $w$ and so this
includes the special point $(a,b)=(2,2)$. The large $w$ asymptotics
for $\kappa$ for the remaining special points are calculated by expanding
in inverse powers of $w$. They are
\begin{itemize}
\item when $a=b=1$ 
  \begin{equation}
%    z_c = \frac{1}{4} + \frac{\pi^2}{4 w^2} -
%    \frac{\pi^2}{w^3} + \frac{\pi^4 + 18\pi^2}{6 w^4} + O(w^{-5}).
%    z_c = \frac{1}{2}+\frac{\pi^2}{4w^2}-\frac{\pi^2}{w^3}
%    + \frac{5\pi^4+144\pi^2}{48w^4} + O(w^{-5})
%\end{equation}
%and so
%\begin{equation}
  \kappa(w,1,1)
  =\log(2)-\frac{\pi^2}{2w^2}+\frac{2\pi^2}{w^3}-
  \frac{72\pi^2+\pi^4}{12w^4} + O(w^{-5}), 
\end{equation}
\item and when $(a,b)=(1,2)$ or $(2,1)$ 
  \begin{equation}
%    z_c = \frac{1}{4} + \frac{\pi^2}{16 w^2} -
%    \frac{\pi^2}{8 w^3} + \frac{\pi^4 + 18\pi^2}{96 w^4} + O(w^{-5}).
%    z_c = \frac{1}{2}+\frac{\pi^2}{16w^2}-\frac{\pi^2}{8w^3}
%    + \frac{5\pi^4+144\pi^2}{768w^4} + O(w^{-5})
%  \end{equation}
%and so
%\begin{equation}
\fl \quad
\kappa(w,1,2) = \kappa(w,2,1) = \log(2) - \frac{\pi^2}{8w^2} +
  \frac{\pi^2}{4w^3} - \frac{72\pi^2+\pi^4}{192w^4} + O(w^{-5}).
\end{equation}
\end{itemize}

\section{Asymptotics for large widths at general parameter values}
\label{asymptotics}

As we have noticed the large $w$ limit is dependent on the order of
limits $w\rightarrow \infty$ and $n\rightarrow \infty$. As such we can
now consider the large $w$ asymptotics of the free energy (which
itself is first defined via an $n\rightarrow \infty$ limit). As
mentioned above, we are interested in the forces between the walls for
large $w$ which also involves large $w$ asymptotics. However, we are
unable to find a closed form solution for $\kappa(w;a,b)$ for arbitrary
$a,b$ and $w$. We can, however, find asymptotic expressions for $\kappa$
for arbitrary $a,b$ and large $w$ without the need for such an
expression.

For general $a$ and $b$ the analysis depends on the region of
parameter space and the plane naturally breaks up into various
regions. These will become different phases when considering the limit
$w \rightarrow \infty$. The analysis proceeds in a self-consistent
manner perturbing around the solution obtained from the special points
in descending powers of $w$.

For $a,b<2$ we can find the asymptotics of $q_c$ as a
function of $w$,
  \begin{equation}
    q_c = \exp\left( 2\pi i \Big/
      \left(w+\frac{2(a+b-ab)}{(2-a)(2-b)}\right)  + O(w^{-4}) \right)  ,
  \end{equation}
  which, mapping back to the $z$ variable, gives
  \begin{equation}
%    z_c = \frac{1}{4} + \frac{\pi^2 }{4 w^2}
%    +\frac{\pi^2 (ab-a-b)}{w^3(2-a)(2-b)}
%    +\frac{\pi^4 k^4}{6 w^4} + \frac{3 \pi^2 k^2 (ab-a-b)^2}{w^4(2-a)^2(2-b)^2}
%    + O(w^{-5})
\fl \quad
z_c = \frac{1}{2} + \frac{\pi^2 }{4 w^2}
    +\frac{\pi^2 (ab-a-b)}{(2-a)(2-b) w^3}
    + \left( \frac{5\pi^4 }{48 } + 
      \frac{3 \pi^2 (ab-a-b)^2}{(2-a)^2(2-b)^2} 
    \right) \frac{1}{w^4} + O(w^{-5})
  \end{equation}
and so
\begin{equation}
\fl \quad
  \kappa = \log(2) - \frac{\pi^2}{2 w^2}  
  - \frac{2\pi^2(ab-a-b)}{(2-a)(2-b) w^3}
  - \left(
    \frac{\pi^4}{12} + \frac{6(ab-a-b)^2}{(2-a)^2(2-b)^2}
    \right) \frac{1}{w^4} + O(w^{-5}).
\end{equation}
  For $b=2$ and $a<2$ we have
  \begin{equation}
    q_c = \exp\left( \pi i \Big/
      \left(w+\frac{a}{2-a}\right) + O(w^{-4})  \right) 
  \end{equation}
  which gives
  \begin{equation}
%    z_c = \frac{1}{4} + \frac{\pi^2}{16 w^2}
%    - \frac{\pi^2 a}{8 w^3(2-a)} + \frac{\pi^4}{96 w^4} 
%    + \frac{3 \pi^2 a^2}{16 w^4(2-a)^2} + O(w^{-5})
\fl \qquad
z_c = \frac{1}{2} + \frac{\pi^2}{16 w^2}
    - \frac{\pi^2 a}{8 (2-a) w^3} + \left( \frac{5 \pi^4}{768} 
    + \frac{3 \pi^2 a^2}{16 (2-a)^2} \right) \frac{1}{w^4} + O(w^{-5})
  \end{equation}
and so
\begin{equation}
\fl \qquad
\kappa = \log(2) - \frac{\pi^2}{8w^2} +  \frac{\pi^2 a}{4(2-a)w^3}
  -\left( \frac{\pi^4}{192} + \frac{3 \pi^2 a^2}{8 (2-a)^2} \right)\frac{1}{w^4} +O(w^{-5}).
\end{equation}
  When $a=2$ and $b<2$ we obtain a similar expression where $b$
replaces $a$.

When $a$ or $b$ is greater than $2$ the asymptotic form changes.
In particular the right-hand side of equation~(\ref{q-sing-eqn}) is
satisfied by either $q = 1/(a-1)$ or $q=1/(b-1)$ and in the limit
as $w \to \infty$ the left-hand side of the equation is zero. Hence we
expand about the point $q = 1/(a-1)$ and find that the next term in
the expansion is exponential in $w$. More precisely, when $a>b$ and
$a>2$ the asymptotic expansion of $q_c(w)$ as $w \to \infty$ is
  \begin{equation}
  \fl \qquad
    q_c = \left(\frac{1}{a-1}\right) -
    \frac{a(a-2)(ab-a-b)}{(a-1)^2(a-b)} \left( \frac{1}{a-1} \right)^w
    + O\left(w \left(\frac{1}{a-1}\right)^{2w} \right).
  \end{equation}
Transforming this back to the $z$ variable gives
  \begin{equation}
  \fl \qquad
    z_c = \frac{\sqrt{a-1}}{a}-\frac{(a-2)^2(ab-a-b)}{2 a (a-b)
      \sqrt{a-1}} \left( \frac{1}{a-1} \right)^w
    + O\left(w \left(\frac{1}{a-1}\right)^{2w} \right)
  \end{equation}
and so the free energy is 
\begin{equation}
  \fl \qquad
  \kappa = \log \left( \frac{a}{\sqrt{a-1}} \right) +
  \frac{(a-2)^2(ab-a-b)}{2(a-1)(a-b)}\left( \frac{1}{a-1} \right)^{w}
    + O\left(w \left(\frac{1}{a-1}\right)^{2w} \right).
\end{equation}

For $b>a$ and $b>2$ one simply interchanges $a$ and $b$ in the above
expressions to obtain the correct results.

Finally, when $a=b>2$ the asymptotic expansion of $q_c(w)$ is
  \begin{equation}
    q_c = \left(\frac{1}{a-1}\right) -
    \frac{a(a-2)}{(a-1)^2} \left( \frac{1}{a-1} \right)^{w/2}
    + O\left(w \left(\frac{1}{a-1}\right)^{w} \right)
  \end{equation}
 and transforming this back to the $z$ variable gives
  \begin{equation}
    z_c = \frac{\sqrt{a-1}}{a}-\frac{(a-2)^2}{2 a \sqrt{a-1}} 
    \left( \frac{1}{a-1} \right)^{w/2}
    + O\left(w \left(\frac{1}{a-1}\right)^{w} \right)
\end{equation}
and
\begin{equation}
  \kappa = \log \left( \frac{a}{\sqrt{a-1}} \right) +
  \frac{(a-2)^2}{2(a-1)}\left( \frac{1}{a-1} \right)^{w/2}
    + O\left(w \left(\frac{1}{a-1}\right)^{w} \right).
\end{equation}

\section{The infinitely wide slit}
\begin{figure}[ht!]
\begin{center}
\includegraphics[width=10cm]{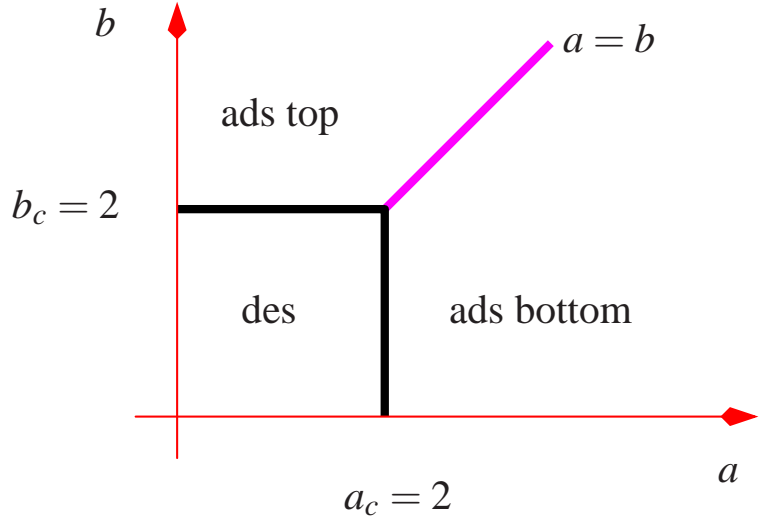}
\caption{\it  Phase diagram of the infinite strip. There are 3 phases: 
desorbed (des), adsorbed onto the bottom wall (ads bottom) and
adsorbed onto the top (ads top). }
\label{phase-diagram} 
\end{center}
\end{figure}

We now come to consider the limit of large slits for infinite
length polymers. Even though there isn't a closed form solution for
finite widths of the equation  (\ref{q-sing-eqn}) we can simply use
the large $w$ asymptotics above to deduce that
%
%one can still deduce 
%the infinite $w$ limit solution for $q_c$ at all parameter
%values. This can be done by considering the large $w$ asymptotics and
%then looking at the dominant term.
%In the next section we will deduce the large $w$ asymptotics from
%which we find 
\begin{equation}
\fl \qquad
        \kappa^{inf-slit}(a,b) \equiv \lim_{w \rightarrow \infty} \kappa(w;a,b)
=\left\{ \begin{array}{ll} \log(2) & \mbox{ if } a,b
        \leq 2 \\ 
\log\left(\frac{a}{\sqrt{a-1}}\right) & \mbox{ if } a > 2 \mbox{ and }
        a>b \\ 
\log\left(\frac{b}{\sqrt{b-1}}\right) & \mbox{ otherwise.} \end{array}
\right. 
\label{inf-strip-free-energy}
\end{equation}
One immediately sees that the free energy depends on both $a$ and
$b$ rather than only on $a$ which reflects the difference 
from the half-plane result (equation \ref{halfplanefe}) noted
earlier. In fact, the
free energy is symmetric in $a$ and $b$ which reflects the observation
that for infinitely long walks the end-points are irrelevant.
For example, as in the case we considered earlier in figure
\ref{kappa-b=3} with $b>2$ the free energy for $a\leq b$ is
$\log(\frac{b}{\sqrt{b-1}})$ (which is approximately $0.752$ for $b=3$) while
it is $\log(\frac{a}{\sqrt{a-1}})$ for $a> b$.  The half-plane phase
diagram contains a desorbed phase for $a<2$ and an adsorbed phase for
$a>2$. The transition between them is second order with a jump in the
specific heat. 
%Hence a naive phase diagram for the infinite strip
%based on the half-plane result does not hold. 
Using (\ref{inf-strip-free-energy}) one can deduce for the infinite
slit that there are 3 phases:
one where the polymer is desorbed for $a,b < 2$; a phase where the
polymer is adsorbed onto the bottom wall for $a > 2$ with $a>b$; and a
phase where the polymer is adsorbed onto the top wall for $b > 2$ with
$b>a$.  This is illustrated in figure
\ref{phase-diagram}.
The low temperature adsorbed phases are characterised by the order
parameter of the thermodynamic density of visits to the bottom and top
walls respectively. There are 3 phase transition lines. The first two are
given by $b=2$ for $0\leq a \leq 2$ and $a=2$ for $0\leq b \leq
2$. These lines separate the desorbed phase from the adsorbed phases
and are lines of second order transitions of the same nature as the
one found in the half-plane model. There is also a first order
transition for $a=b >2$ where the density of visits to each of the
walls jumps discontinuously  on crossing the boundary non-tangentially.

%\begin{figure}[ht!]
%\begin{center}
%\includegraphics[width=8cm]{figures/pd_comp1.eps}
%\caption{\it  naive phase diagram}
%\label{nphasediagram} 
%\end{center}
%\end{figure}
%
%\begin{figure}[ht!]
%\begin{center}
%\includegraphics[width=8cm]{figures/zc_b_a_schem.eps}
%\caption{\it  free energy for $b>2$}
%\label{feb3} 
%\end{center}
%\end{figure}
%
%\begin{figure}[ht!]
%\begin{center}
%\includegraphics[width=10cm]{figures/pd_comp2.eps}
%\caption{\it  phase diagram}
%\label{phasediagram} 
%\end{center}
%\end{figure}

\section{Forces between the walls.}

%The free energy of the system is $\kappa(w) = -\log z_c(w)$. 
Let us define the effective force between the walls induced by the
polymer as
%Since the
%width is a discrete variable the force induced on the system would be
%calculated as
%\begin{equation}
%  \bar{\mathcal{F}}(w) = \kappa(w+1) -  \kappa(w)
%\end{equation}
%However, as we are interested in the large width case we will rather
%consider the normal continuous force
\begin{equation}
  \mathcal{F}(w) = \diff{\kappa(w)}{w} 
  =\frac{1}{z_c(w)}\frac{-\partial z_c(w)}{\partial w}.
\end{equation}
Before examining the general case for large $w$ it is worth
considering the general case for small $w$ numerically and at the
special points for all $w$ where it can be found exactly. 
%
%\begin{figure}[ht!]
%\begin{center}
%\includegraphics[width=10cm]{figures/zcw_1_1.eps}
%\caption{\it  free energy vs $w$ for $a=b=1$}
%\label{force1} 
%\end{center}
%\end{figure}
%\begin{figure}[ht!]
%\begin{center}
%\includegraphics[width=10cm]{figures/zcw_4_4.eps}
%\caption{\it  free energy vs $w$ for $a=b=4$}
%\label{force4} 
%\end{center}
%\end{figure}
To begin, at the special points we can calculate the induced force exactly
since we know the value of $\kappa(w)$ exactly. We note that since
$\kappa(w)$ is monotonic at each of these points the induced force has
one sign for all $w$. In particular,
  \begin{itemize}
  \item when $a=b=1$ then $1/z_c = 2 \cos(\pi/(w+2))$ and so
    \begin{equation}
      \mathcal{F} = \frac{\pi }{(w+2)^2} \tan(\pi/(w+2))
      = \frac{\pi^2}{w^3} - \frac{6 \pi^2}{w^4} + O(w^{-5})
    \end{equation}
    which is positive and hence repulsive.
  \item When $(a,b)=(2,1), (1,2)$ then $1/z_c = 2 \cos(\pi/2 (w+1))$
    and so
    \begin{equation}
      \mathcal{F} = \frac{\pi}{2 (w+1)^2} \tan(\pi/ 2 (w+1))
      = \frac{\pi^2}{4 w^3} - \frac{3 \pi^2}{4w^4} + O(w^{-5})
    \end{equation}
    which is positive and hence repulsive.
%  \item when $a=b=2$ then $1/z_c = 2$ so 
%    \begin{equation} 
%      \mathcal{F} = 0
%    \end{equation}
  \item When $ab=a+b$ including $a=b=2$ then $1/z_c =
    \frac{\sqrt{a-1}}{a} = \frac{\sqrt{b-1}}{b}$ and so 
    \begin{equation}
      \mathcal{F} = 0.
    \end{equation}
    \end{itemize}
    If one considers more general values of $a$ and $b$ numerically
    one observes that in each region of the phase plane $\kappa(w)$ is
    monotonic and hence the force takes on a unique sign at each value
    of $a$ and $b$.
%\subsection{General parameter asymptotics}
At general $(a,b)$ we are able to find the asymptotic force for large
$w$ by using the large $w$ asymptotic expression for $\kappa(w;a,b)$ 
which we calculated above.

\begin{figure}[ht!]
\begin{center}
\includegraphics[width=10cm]{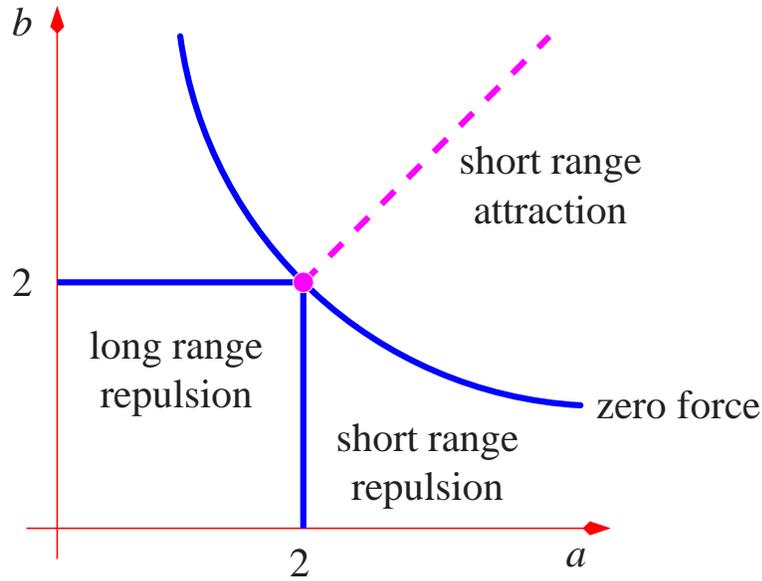}
\caption{\it  A diagram of the regions of different types of effective 
force between the walls of a slit. Short range behaviour refers to
exponential decay of the force with slit width while long range refers 
to a power law decay. The zero force curve is the special
curve given by $ab=a+b$. On the dashed line there is a singular change 
of behaviour of the force. }
\label{force-diagram} 
\end{center}
\end{figure}

Using the asymptotic expressions for $\kappa$ found in section
\ref{asymptotics} in the $(a,b)$-plane we obtain the asymptotics for
the force.
  \begin{itemize}
  \item For $a,b<2$
    \begin{equation}
\mathcal{F}=\frac{\pi^2}{w^3} +\frac{6\pi^2(ab-a-b)}{(2-a)(2-b)w^4} + O(w^{-5})
    \end{equation}
    which is positive and hence repulsive.
  \item For $b=2,a<2$
    \begin{equation}
\mathcal{F}=\frac{\pi^2}{4 w^3} - \frac{3 \pi^2 a}{4 (2-a) w^4} + O(w^{-5})
    \end{equation}
    which is positive and hence repulsive.
  \item For $a>b$ and $a>2$
    \begin{equation}
    \fl \qquad
\mathcal{F} = - \frac{(a-2)^2(ab-a-b) \log (a-1)}{2 (a-1)(a-b)} \left(\frac{1}{a-1}\right)^w
      + O\left(\left(\frac{1}{a-1}\right)^{2w} \right).
    \end{equation}
%    Note that the force changes sign when $(a,b)$ crosses the curve
%    $ab=a+b$ --- to the north-east of the curve the force is negative
%    (and hence attractive) and to the south-west it is positive (and
%    hence repulsive).
For $b>a$ and $b>2$ one simply interchanges $a$ and $b$ in the above
expressions to obtain the required result. 
  \item At $b=a>2$
    \begin{equation}
      \mathcal{F} = - \frac{(a-2)^2 \log(a-1) }{4 (a-1) } \left(\frac{1}{a-1}\right)^{w/2}
      + O\left(\left(\frac{1}{a-1}\right)^{w} \right).
    \end{equation}
    This is negative for all $a>2$ and so the induced force is attractive.
  \end{itemize}
The regions of the plane which gave different asymptotic expressions
for $\kappa$ and hence different phases for the infinite slit clearly
also give different force behaviours. For the square $0\leq a,b\leq 2$
the force is repulsive and decays as a power law (ie \ it is long-ranged) 
while outside this square the force decays exponentially and so is
\emph{short-ranged}. This change coincides with the phase boundary of
the infinite slit phase diagram. However, the special curve $ab=a+b$
is a line of zero force across which the force, while short-ranged on
either side (except at $(a,b)=(2,2)$), changes sign. Hence this curve
separates regions where the force is attractive (to the right of the
curve) and repulsive to the left of the curve. The line $a=b$ for
$a>2$ is also special and, while the force is always
short-ranged and attractive, the range of the force on the line is
discontinuous and twice the size on this line than close by. All these
features lead us to us a \emph{force diagram} that encapsulates these
features. This diagram is given in figure \ref{force-diagram}.

\section{Discussion}

We have solved and analysed in the limit of infinite length three types of
polymer configuration, being loops, bridges and tails, in a
two-dimensional slit geometry. The types of configurations considered
are directed walks based on Dyck paths and the walks interact with
both walls of the slit. We have found the exact generating functions
at arbitrary width. We have calculated the free energy exactly at
various points for arbitrary width, and, importantly, asymptotically
(for large widths) for arbitrary $a$ and $b$. This has allowed us to
map out the phase diagram for infinite slits: for finite width the
free energy is an analytic function of the Boltzmann weights $a$ and
$b$. This phase diagram is \emph{different} to that obtained in the
half-plane even though the generating functions for the half-plane are
a formal limit of the finite width generating functions. This arises
because the order of the infinite width and infinite polymer length
limits (both of which are needed to see a phase transition) are \emph{not}
interchangeable.

From the large width asymptotics we have mapped regions of the plane
where the free energy decreases or increases with increasing width
and, also, whether this happens with an algebraic decay or exponential
decay. Using this we have delineated regions where the induced force
between the walls is short-ranged or long-ranged and whether it is
attractive or repulsive. Three types of behaviour occur: long-ranged
repulsive, short-ranged repulsive and short-ranged attractive.
Applying such a model to colloidal dispersions implies that the
regions where the force is long-ranged and repulsive support steric
stabilisation while the regions where the force is short-ranged and
attractive promote sensitized flocculation.

Given the curious interchange of limits phenomenon and from the point
of view of the application of this theory to colloidal dispersions it
would be interesting to investigate the finite polymer length cases
both analytically and numerically with a view of searching for a scaling
theory for the crossover from half-plane to slit type behaviour.

The model considered here is a directed walk in two 
dimensions.  Although we expect the general features 
(such as the phase diagrams sketched in Figures 
7 and 8) to be the same for self-avoiding walk models 
in both two and three dimensions, such models are beyond
the reach of analytic treatments.  It would be interesting
to investigate the detailed behaviour of such models by Monte Carlo 
methods or by exact enumeration coupled with series analysis
techniques.

\section*{Acknowledgements} 
%***************************************************************
 Financial support from  the Australian Research Council is gratefully
acknowledged by RB, ALO and AR. Financial support from NSERC 
of Canada is gratefully acknowledged by SGW.
%***************************************************************

%\pagebreak
\begin{sloppypar}
\section*{References}

\vspace{0.1in} \noindent
Andre, D 1887 \emph{C. R. Acad. Sci. Paris}
\textbf{105} {436--437}

\vspace{0.1in} \noindent
Abramowitz M and Stegun I A (Eds.). \emph{Orthogonal Polynomials.
  Ch.~22 in Handbook of Mathematical Functions with Formulas, Graphs,
  and Mathematical Tables, 9th printing}.  New York: Dover, 771--802, 1972.

\vspace{0.1in} \noindent
Bertrand J 1887 \emph{C. R. Acad. Sci. Paris}
\textbf{105} {369}

\vspace{0.1in} \noindent
Bousquet-M\'elou M and Rechnitzer A 2002 \emph{Discrete Math.}
\textbf{258} {235--274}

\vspace{0.1in} \noindent
Brak R,   Essam J W and Owczarek A L 1998   \emph{J. Stat. Phys.} {\bf 93}  {155--192} 

\vspace{0.1in} \noindent
Brak R,   Essam J W and Owczarek A L 1999   \emph{J. Phys. A.} {\bf 32}  {2921--2929} 

\vspace{0.1in} \noindent        
 Brak R and   Essam J  W 2001 \emph{J. Phys. A.} {\bf 34}  {10763--10781}

\vspace{0.1in} \noindent
DiMarzio E A and Rubin R J 1971 \emph{J. Chem. Phys.}  {\bf 55} 4318--4336

\vspace{0.1in} \noindent Deutsch E  1999 \emph{Discrete Math.}  {\bf
  204} {167--202}

\vspace{0.1in} \noindent Flajolet P 1980 \emph{Discrete Math.}  {\bf
  32} {125--161}
        
\vspace{0.1in} \noindent Hammersley J M and Whittington S G 1985
\emph{J. Phys. A: Math. Gen.}  {\bf 18} 101--111

\vspace{0.1in} \noindent Janse van Rensburg E J, 2000 \emph{The
  Statistical Mechanics of Interacting Walks, Polygons, Animals and
  Vesicles}, Oxford University Press, Oxford.

\vspace{0.1in} \noindent Middlemiss K M, Torrie G M and Whittington S
G 1977 \emph{J. Chem. Phys.}  {\bf 66} 3227--3232

\vspace{0.1in} \noindent Stanton D and White D, 1986 \emph{Constructive Combinatorics}, Springer-Verlag.

\vspace{0.1in} \noindent Szeg\"{o} G. \emph{Orthogonal Polynomials,
  4th ed}. Providence, RI: Amer. Math. Soc., 44--47 and 54--55, 1975.

\vspace{0.1in} \noindent Viennot G 1985 \emph{Lecture Notes in
  Mathematics} {\bf 1171} 139--157

\vspace{0.1in} \noindent Wall F T, Seitz W A, Chin J C and de Gennes P
G 1978 \emph{Proc. Nat. Acad. Sci.} {\bf 75} 2069--2070

\vspace{0.1in} \noindent Wall F T, Seitz W A, Chin J C and Mandel F
1977 \emph{J. Chem. Phys.}  {\bf 67} 434--438

\end{sloppypar}

 \end{document}